\documentclass[%
 reprint,
showpacs,%preprintnumbers,
 amsmath,amssymb,
 aps,
pre,
floatfix,
]{revtex4-1}

\usepackage{graphicx}% Include figure files
\usepackage{dcolumn}% Align table columns on decimal point
\usepackage{bm}% bold math
\begin{document}

%\preprint{}

\title{Mutual Information Rate-Based Networks in Financial Markets}% Force line breaks with \\
%\thanks{A footnote to the article title}%

\author{Pawe\l{} Fiedor}
\email{s801dok@wizard.uek.krakow.pl}
%\author{Artur Ho\l{}da}%
%\email{aholda@uek.krakow.pl}
\affiliation{%
 Cracow University of Economics\\
 Rakowicka 27, 31-510 Krak\'{o}w, Poland
}%

\date{\today}

\begin{abstract}
In the last years efforts in econophysics have been shifted to study how network theory can facilitate understanding of complex financial markets. Main part of these efforts is the study of correlation-based hierarchical networks. This is somewhat surprising as the underlying assumptions of research looking at financial markets is that they behave chaotically. In fact it's common for econophysicists to estimate maximal Lyapunov exponent for log returns of a given financial asset to confirm that prices behave chaotically. Chaotic behaviour is only displayed by dynamical systems which are either non-linear or infinite-dimensional. Therefore it seems that non-linearity is an important part of financial markets, which is proved by numerous studies confirming financial markets display significant non-linear behaviour, yet network theory is used to study them using almost exclusively correlations and partial correlations, which are inherently dealing with linear dependencies only. In this paper we introduce a way to incorporate non-linear dynamics and dependencies into hierarchical networks to study financial markets using mutual information and its dynamical extension: the mutual information rate. We estimate it using multidimensional Lempel-Ziv complexity and then convert it into an Euclidean metric in order to find appropriate topological structure of networks modelling financial markets. We show that this approach leads to different results than correlation-based approach used in most studies, on the basis of 15 biggest companies listed on Warsaw Stock Exchange in the period of 2009-2012 and 91 companies listed on NYSE100 between 2003 and 2013, using minimal spanning trees and planar maximally filtered graphs.
\end{abstract}

% PACS, the Physics and Astronomy
                             % Classification Scheme.
%\keywords{Suggested keywords}%Use showkeys class option if keyword
                              %display desired
\maketitle

%\tableofcontents

\section{Introduction}

Financial markets treated as complex systems are well-defined. Nonetheless economists lack a fundamental theory behind their complex behaviour, which opened doors for other scientists, such as mathematicians and physicists, to study those systems. The lack of theory leads to an assumption that the time series describing stock returns (usually logarithms of those, also changes in prices of any other assets such as market indices and currency rates) are unpredictable \cite{Samuelson:1965}. Within this paradigm the evolution of stock prices can only be explained by random processes. But if we assume that price formation is a stochastic process then it's natural to ask whether these processes are independent for different financial instruments or whether there exist common economic factors driving the price formation processes for numerous financial instruments. Common economic factors were not found in scientific research to date, nonetheless tools and procedures developed first to model physical systems \cite{Mandelbrot:1963,Kadanoff:1971,Mantegna:1991} are used to characterise the interdependencies of different financial instruments, or in other words to classify the financial instruments according to their interdependencies.

Classification of any data is very important in science, and particularly so in fields with vast amounts of data such as statistical finance. Classification allows easier and more effective understanding and learning \cite{Jain:1988}. Classification can be exclusive or overlapping, and also supervised or unsupervised. Studies in financial markets use exclusive unsupervised classifications. The procedure of obtaining such classifications is called clustering.

Clustering proceeds over a set of objects which are to be classified according to a set of properties (called the characteristic vector) assigned to each object. Applying clustering separates the objects into groups called clusters based merely on the characteristic vectors themselves. Clustering most often relies only on certain parts of the characteristic vectors and not the entirety of these. Therefore finding the relevant characteristic plays a major role in the process of clustering. Clustering organises objects as a single grouping of individuals into non-overlapping clusters or as a hierarchy of nested partitions. The former is known as partitional clustering, whereas the latter is called hierarchical clustering. Hierarchical clustering is most often used in financial market analysis due its ability to create a dendrogram, which makes it easier to analyse the clustering of financial instruments into groups visually and intuitively \cite{Hubert:1976}. Importantly any hierarchical clustering classification can be converted into partitional clustering through horizontal flattening of the dendrogram, but the reverse is not true, partitional clustering does not retain all the information of hierarchical clustering, therefore one is unable to move from partitional to hierarchical clustering effortlessly. Clustering methods are widely used in a variety of applications and many techniques have been developed \cite{Jain:1988}.

As hinted in the above description the crucial problem in any clustering procedure is the choice of the measure of proximity between objects. A measure of this sort obviously has to be obtained from the characteristic vectors. It's usually a measurement of similarity (or equivalently dissimilarity). Any valid measure of similarity will not be an Euclidean metric and therefore needs to be manipulated for the sake of easiness of use in many applications, while a measure of dissimilarity usually satisfies the standard axioms of an Euclidean metric (positivity, symmetry, and triangle inequality). A matrix of all pairwise proximity measures is called the proximity matrix (similarly as a matrix of all pairwise correlations is called the correlation matrix). Having such measure hierarchical clustering methods can then either use them only at the first level of the hierarchy and for all other levels derive the proximities between clusters from the proximities of their elements, or alternatively the methods can calculate them at every level from the original measures \cite{Kraskov:2005}. Methods in the latter group allow more flexibility, but are more also computationally expensive. This is in most cases not a problem, and thus those are usually used in analysing financial markets.

For various applications many similarity measures have been developed and implemented, but in analysing the financial markets the researchers are persistently using only Pearson's correlation coefficient and its derivatives. While correlation-based graphs constitute a powerful tool for detecting and analysing (also visually) parts of the most statistically robust information present in the correlation-based characteristic vectors \cite{Tumminello:2010} it is nonetheless troubling as will become clear in the course of this introduction. The correlation structure of log returns of financial instruments (most often stocks, but also indexes and foreign exchange rates) contains key information for many practical applications such as portfolio optimisation, risk management, and option pricing \cite{Mantegna:1999}. Such correlation structures have been investigated for time series describing stock returns \cite{Laloux:1999,Plerou:1999,Mantegna:1999,Tumminello:2010,Akemann:2011}, market index returns \cite{Bonanno:2000,Maslov:2001,Drozdz:2001,Coelho:2007,Gilmore:2008,Eryigit:2009,Song:2011,Sandoval:2012} and currency exchange rates \cite{McDonald:2005}. The tools for analysing such correlation structure contain spectral density analysis of the eigenvalues of the correlation matrix, tools of multivariate analysis, and random matrix theory \cite{Laloux:1999,Plerou:1999,Akemann:2011}. Similarity based graphs, or in other words networks associated with the similarity matrices \cite{Mantegna:1999,Onnela:2002,Onnela:2003,Tumminello:2005,Tumminello:2007,Coronnello:2007,Tumminello:2010}, are also used. In all cases the point is to extract the most relevant information present in the similarity matrix.

The insistence of researchers to use Pearson's correlation coefficient or related measures as proximity measure in hierarchical clustering is surprising. It is well-known since the 1990s that financial markets, and particularly time series describing returns on financial instruments, are involving terms that are not of the first degree. In fact the interest in non-linear dynamics in financial markets has first strongly emerged after the stock market crash of October 19, 1987 \cite{Hsieh:1991}. Frank and Stengos studied the rates of return on commodities (particularly gold and silver) and concluded that there exists evidence of non-linear deterministic price formation process \cite{Frank:1989}. D. Hsieh studied daily currency exchange rate changes for five major currencies and found evidence for the presence of substantial non-linearity in a multiplicative form \cite{Hsieh:1989}. Scheinkman and LeBaron have also found evidence indicating the presence of non-linear dependence in weekly log returns for financial indices \cite{Scheinkman:1989}. In 1991 the first book devoted to the non-linear dynamics of financial markets has been published \cite{Brock:1991}. In 1995 Abhyankar, Copeland and Wong tested intra-day data from FTSE-100 index for the presence of non-linear dependence and indeed found evidence of such \cite{Wong:1995}. There is now overwhelming evidence of non-linear dynamics in stock returns \cite{Brock:1991,Qi:1999,McMillan:2001,Sornette:2002,Kim:2002}, market index returns \cite{Franses:1996,Wong:1995,Chen:1996,Wong:1997,Ammermann:2003}, and currency exchange rate changes \cite{Hsieh:1989,Brock:1991,Rose:1991,Brooks:1996,Wu:2003}. Therefore the assumptions that only linear dependencies are relevant in financial markets found in hierarchical clustering methodology used in econophysics is baffling. In this paper we propose to amend the methodology of clustering for financial data so that the measure of similarity takes non-linear dependencies into account.

As stated above Pearson's correlation coefficient is strictly not sensitive to any non-linear dependencies. Therefore such analysis can potentially miss important features of any dynamical system, particularly financial systems which have been shown to present significantly non-linear behaviour. Correlation coefficient is then contrasted by the measure of mutual information (MI), which is differing from correlation due to its information theoretic background \cite{Cover:1991}, which incidentally makes it a much more general measure. In fact $MI = 0$ if and only if the two studied random variable are strictly (statistically) independent. Mutual information is then a natural measure which can be used to extend the similarity measure to make it sensitive to non-linear dependencies, and has indeed been successfully used in some applications \cite{Zhou:2007,Zhou:2009,Muller:2012}. Mutual information is a measure of great importance in many fields precisely because it quantifies both the linear and non-linear interdependencies between two systems or stochastic processes. Mutual information can be interpreted as a measure of how much information two studied systems exchange or two studied stochastic processes or data sets share. Due to these characteristics mutual information is suitable for many applications, and has been used successfully particularly enhance the understanding of the development and functioning of the brain in neuroscience \cite{Sporns:2004,Bialek:1998,Bialek:2002}, to characterise \cite{Donges:2009,Palus:2001} and model various complex and chaotic systems \cite{Fraser:1986,Parlitz:1998,Kantz:2004}, and also to quantify the information capacity of a communication system \cite{Haykin:2001}. Additionally mutual information provides a convenient way to identify the most relevant variables with which to describe the behaviour of a complex system \cite{Rossi:2006}, which is of paramount importance in modelling those systems, and indeed to the methodology of this paper.

The calculation or indeed estimation of mutual information in dynamical systems is met with three important difficulties however \cite{Paninski:2003,Palus:2001}. Mutual information is precisely defined only for random processes without memory. Unfortunately we know that most dynamical systems are not strictly memoryless and indeed financial markets have been shown to contain a degree of memory in their random walk \cite{Floros:2007,Hsieh:2009,Manap:2011}. Secondly, to calculate mutual information it is often necessary to find probabilities of significant events, and defining significant events may not be a trivial issue, as significant events are not always precisely known. Thirdly, data sets and samples have finite size. This prevents the researchers from calculating the probabilities correctly. As a consequence, mutual information can often only be calculated with a bias \cite{Palus:2001,Steuer:2002,Papana:2009}. Nonetheless those restrictions does not make mutual information useless, particularly the third one is true of any other similarity measure (including correlation), and the second one is a matter of careful design of methodology, while the first problem is not severe and can be contained by using methods which are asymptotically precise even for processes with memory.

In this study we propose not to use mutual information itself as a measure of similarity between financial instruments, even though this itself should present a relatively good extension to the correlation-based studies, but instead to calculate the amount of information exchanged between two nodes (or clusters of nodes) in a dynamical network (or between two data sets) per unit of time, or the mutual information rate (MIR), and to use it as the similarity measure for the hierarchical dependency networks. Mutual information is based on the Shannon's concept of entropy, and consequently the dynamical extension of mutual information, that is the mutual information rate, is in turn based on the dynamical extension of entropy or the entropy rate \cite{Shannon:1948}. Therefore to estimate mutual information rate a method of estimating entropy rate is needed beforehand. Having in mind the first of the three problems mentioned above, we know that the classical definition of entropy rate is based on an asymptotic limit \cite{Shannon:1948,Gao:2008}, hence it's not easy to find an accurate estimator for finite-size samples, which is indeed the case in financial markets, especially when considering daily price changes \cite{Navet:2008}. The concept of complexity in the sense of Kolmogorov (complexity of a sequence is the size of the smallest binary program which can produce this sequence \cite{Cover:1991}) can be used to obtain accurate estimates of the entropy rate, that is one fast approaching the real value with the increase of the sample size. Using the implementation of Lempel-Ziv complexity (LZC) \cite{Lempel:1976} allows us to gain the advantage on two finite size issues (third problem mentioned above): firstly the problem of sampling or an accurate control of the statistical fluctuations \cite{Amigo:2006} and secondly a better estimation of an asymptotic quantity \cite{Lesne:2009}. The Lempel-Ziv complexity has been mostly used in neurobiology, and many studies of neural spike trains have been performed using this measure \cite{Amigo:2004,Aboy:2006,Christen:2006}, but it has also been used in a limited number of studies in econophysics \cite{Navet:2008,Fiedor:2013}. Nonetheless most of these studies use one-dimensional analysis, and for estimating mutual information rate a two-dimensional analysis is required. In this paper we use an extension of Lempel-Ziv complexity to multidimensional signals \cite{Zozor:2005,Blanc:2008} to study the estimate of higher order correlations between pairs of financial instruments. The validity of estimating mutual information using Lempel-Ziv complexity for non-linear time series has been confirmed in earlier studies \cite{Amigo:2004,Zhong:2007}, therefore it seems natural that the same should hold true for the dynamical extension of these measures.

The mutual information rate can also be understood as a measurement of all the interdependencies between the spatio-temporal organisation of the observed sequences (realisations of stochastic processes) $X$ and $Y$ and measurement of the degree to which these two sequences produce independent information on the same underlying dynamics of the whole system. As the entropy rate $HR(X)$ measures the temporal structure of the sequence $X$ better than linear statistical measures such as the Pearson's correlation coefficient, the production of mutual information per unit of time (MIR) also provides a more complete quantification of the interrelations between the two sequences $X$ and $Y$ than covariance (mutual information and mutual Lempel-Ziv complexity account for all interdependencies, not only the linear ones). Mutual information rate and mutual Lempel-Ziv complexity also provide a better account of the spatio-temporal structure of the sequences, as these sequences are then not only joint realisations of two random variables, but instead a joint realisation of one random process, rendering the mutual Lempel-Ziv complexity much more meaningful than a collection of quantities computed for singular random variables \cite{Blanc:2008}.

The recent financial crisis renders the investigation into the complex nature of the financial markets and their dynamical properties more important than ever. The complexity of the financial markets and their behaviour in the recent years, together with the very fast dynamics (e.g. the so-called flash crash), means that we no longer can ignore the non-linearity of financial markets without any loss of important information characterising such systems. Therefore in this paper we extend the known methodology of hierarchical clustering of the financial data and creating dependency networks, which present only the most important interdependencies on the studied market in an intuitive way, by exchanging the similarity measure from the Pearson's correlation coefficient to the information-theoretic approach using mutual information rate estimated by multidimensional Lempel-Ziv complexity. We then apply it to log returns on Warsaw's Stock Exchange and NY Stock Exchange in order to show how different the results are from the ones obtained using correlation coefficient.

\section{Similarity measure}

The topological arrangement of the nodes in network-based model (particularly when the financial market is being studied) is most often based on the the Pearson's correlation coefficient, and in the case of financial markets the correlation is taken of the difference of logarithms of closing prices for two consecutive days. Such correlation coefficient is estimated for all pairs of financial instruments in the studied system. The Pearson's correlation coefficient mentioned above is defined as \cite{Feller:1971}:
\begin{equation}
\rho_{X,Y}=\frac{E(XY) - E(X)E(Y)}{\sqrt{(E(X^2)-E(X)^2)(E(Y^2)-E(Y)^2)}}
\end{equation}
where $X$ and $Y$ are the log price change stochastic processes for two studied financial instruments. The correlation coefficient is estimated for a given period.

The properties of correlation require that the correlation matrix is symmetric, with the diagonal being filled with $\rho_{X,X}=1$. It follows then that such matrix contains only $n~(n-1)/2$ meaningful correlation coefficients \cite{Mantegna:1997}. As stated above the Pearson's correlation coefficient being a similarity measure is not an Euclidean metric, therefore it can't be used directly to determine the network topology. Thus there is a need to form a generalised metric based on correlation, to find an approximate distance between the nodes in a network. Usually the below is used:
\begin{equation}
\delta(X,Y)=1-\rho_{X,Y}^2.
\end{equation}

This form guarantees that $\delta(X,Y)$ is an Euclidean metric, that is it conforms to the three axioms:
\begin{enumerate}
\item $\delta(X,Y)=0$\; \text{if and only if}\; $X=Y$;
\item $\delta(X,Y)=\delta(Y,X)$;
\item $\delta(X,Y) \le \delta(X,Z) +\delta(Z,Y)$.
\end{enumerate}

To extend such measure to include non-linear dependencies we propose to base the topological arrangement of the nodes in a network on the mutual information rate between closing prices for two consecutive days for two financial assets. To define mutual information rate we shall first discuss Shannon's formulation of entropy, entropy rate and mutual information \cite{Shannon:1948}. Entropy rate is a term derivative to the notion of entropy, which measures the amount of uncertainty in a random variable. The Shannon's entropy of a single random variable $X$ is defined as 
\begin{equation}
	\label{eq:Def_entropy}
H(X) = -\sum_{i} p(x_i) \log_2 p(x_i) 
\end{equation}
summed over all possible outcomes $\{x_i\}$ with respective probabilities of $p(x_i)$ \cite{Shannon:1948}. 
For two random variables $(X,Y)$, joint entropy $H(X,Y)$ measuring the uncertainty associated with both, and conditional entropy $H(X|Y)$ measuring uncertainty in one random variable assuming the other has been observed, can be calculated. The joint entropy and conditional entropy are related in a following manner: 
\begin{equation}
	\label{eq:Ent_joint_cond}
H(X|Y) = H(X,Y) - H(Y)
\end{equation}

Shannon also introduced the entropy rate, which generalises the notion of entropy for sequences of dependent random variables. For a stationary stochastic process $X = \{X_i\}$, the entropy rate is defined as:
\begin{equation}
	\label{eq:Def_entropy_rate}
	HR(X) = \lim_{n \rightarrow \infty} \frac{1}{n} H(X_1, X_2, \dots, X_n)
\end{equation}
\begin{equation}
	\label{eq:Def_entropy_rate_2}
	HR(X) = \lim_{n \rightarrow \infty} H(X_n | X_1, X_2, \dots, X_{n-1})
\end{equation}
where Eq. \eqref{eq:Def_entropy_rate} holds true for all stochastic processes, but Eq. \eqref{eq:Def_entropy_rate_2} requires stationarity of the process.

We can therefore interpret entropy rate as a measure of the average uncertainty left in the generation of information in a process at time $n$ having observed the complete history up to that point. Theory of information defines entropy rate of a stochastic process as the amount of new information created in a unit of time \cite{Cover:1991}. Joint and conditional entropy rates can similarly be defined and interpreted.

Based on the concept of entropy we can also define mutual information, which has been proposed by Shannon in the following way \cite{Shannon:1948}. Given two discrete random variables $X$ and $Y$ mutual information between them is defined as:
\begin{equation}
I_S(X,Y) = \sum_{y\in{}Y}\sum_{x\in{}X}p(x,y)\log{\frac{p(x,y)}{p(x)p(y)}},
\end{equation}
where $p(x,y)$ is the joint probability distribution function of $X$ and $Y$ and $p(x)$ and $p(y)$ are the marginal probability distributions. For completeness we also define the same for continuous random variables:
\begin{equation}
I_S(X,Y) = \int_{Y}\int_{X}p(x,y)\log{\frac{p(x,y)}{p(x)p(y)}} \, \mathrm{d}x \, \mathrm{d}y,
\end{equation}
where $p(x,y)$ is the joint probability density function of $X$ and $Y$ and $p(x)$ and $p(y)$ are the marginal probability density functions.

Mutual information can be equivalently defined in terms of entropy:
\begin{equation}
I_S(X,Y) = H(X) + H(Y) - H(X,Y),
\label{IS}
\end{equation}
where $H(X)$ and $H(Y)$ are the marginal entropies and $H(X,Y)$ is the joint entropy of $X$ and $Y$. Mutual information measures the amount of information shared by $X$ and $Y$, or in other words how much the information about one stochastic process reduces uncertainty about the other. Mutual information is non-negative and $I_S(X,X)=H(X)$.

The mutual information rate (MIR) was also first introduced by Shannon \cite{Shannon:1948} as the rate of actual transmission \cite{Blanc:2011} and was consequently more rigorously defined by other researchers \cite{Dobrushin:1959,Gray:1980}. Just as entropy rate represents entropy per unit of time, mutual information rate represents the mutual information exchanged between two dynamical variables per unit of time. To simplify the calculation of the MIR, if we have two continuous dynamical variables, we transform them into two discrete symbolic sequences $X$ and $Y$ (we will need discrete variables for the calculations of Lempel-Ziv complexity anyway). For such sequences the mutual information rate is defined by:
\begin{equation}
 MIR=\lim_{n \rightarrow \infty} \frac{I_S(n)}{n}, 
\label{original_MIR}
\end{equation} 
where $I_S(n)$ represents mutual information between the two sequences $X$ and $Y$ calculated by considering words of length $n$.

The mutual information is a fundamental quantity due to its general information theoretic nature, hence using its dynamical extension to quantify the dependencies between financial instruments seems natural. Its maximal value gives the information capacity between the two studied sources of information. Therefore methods of calculating mutual information rate or the bounds of it are of vital importance to many applications. Researchers also showed that it can be reliably estimated with no need for stationarity, statistical stability, or a memoryless source \cite{Verdu:1994}. As mutual information rate is defined over a limit and from probabilities it cannot be easily calculated, especially if it is to be calculated from trajectories in the phase space of a large complex system. In fact these difficulties are similar to the ones found in the calculation of the Kolmogorov-Sinai entropy $H_{KS}$ \cite{Kolmogorov:1959,Sinai:1959}.

The above definitions do not actually present an obvious way to calculate the mutual information rate in practise. One of the ways to do it is to use the Kolmogorov's algorithmic complexity to estimate entropy rate and mutual information rate. But the mutual information rate and its bounds can also be defined in terms of Lyapunov exponents and predictability horizon. In dynamical systems with fast decay of interdependencies the mutual information in Eq. \eqref{IS} measures the amount of information shared between $X$ and $Y$ which is produced within a special time interval $T$, where $T$ represents the time it takes for the dynamical system to lose its memory from the initial state, or for the interdependencies to decay to zero. In other words $T$ is the predictability horizon of this system. Interdependencies in such system cannot be described by linear correlation, but only a non-linear correlation defined in terms of the evolution of spatial probabilities. Therefore, the mutual information rate between the dynamical variables $X$ and $Y$ can be estimated by:
\begin{equation}
MIR=\frac{I_S}{T}
\label{MIR_introduction}
\end{equation}
In chaotic systems (sensitive to initial conditions) predictions are only possible for times smaller than the predictability horizon time $T$, which can be estimated by:
\begin{equation}
  T \approx \frac{1}{\lambda_1}\log{\left[ \frac{1}{\epsilon} \right]}.
\label{T} 
\end{equation}
where $\lambda_1$ is the largest positive Lyapunov exponent measured in space divided into partitions of size $\epsilon$. 

Nonetheless we shall use the first mentioned concept to estimate mutual information rate, that is the Lempel-Ziv complexity, which can be used to estimate both entropy rate and mutual information rate, since it's connected with the complexity in the Kolmogorov sense. In 1965 Kolmogorov defined the complexity of a sequence as the size of the smallest binary program which can produce this sequence \cite{Cover:1991}. This definition is not operational, therefore intermediate measurements are used. Lempel-Ziv algorithm is one of those measurements, which test the randomness of data series. This algorithm has been first introduced by Jacob Ziv and Abraham Lempel in 1977 \cite{Lempel:1977}. On this basis there has been a number of estimators of entropy rate created. In this article we follow \cite{Navet:2008} and use the estimator created by Kontoyiannis in 1998 (estimator $a$) \cite{Kontoyiannis:1998}. This estimator is widely used \cite{Kennel:2005,Navet:2008} and it was shown that it has better statistical properties than previous estimators based on Lempel-Ziv algorithm \cite{Kontoyiannis:1998}, though there is a large choice of slightly different variants to choose from \cite{Gao:2008}, which is largely irrelevant.

Formally to calculate the entropy rate of a random variable $X$, the probability of each possible outcome $p(x_i)$ must be known. When these probabilities are not known, entropy can be estimated by replacing the probabilities with relative frequencies from observed data. The mentioned estimator is defined as:
\begin{equation}
	\label{eq:LZ_complexity}
	\hat{HR}_{lz} = \frac{n \log_2 n}{\sum_i \Lambda_i},
\end{equation}
where $n$ denotes the length of the time series, and $\Lambda_i$ denotes the length of the shortest substring starting from time $i$ that has not yet been observed prior to time $i$, i.e. from time $1$ to $i-1$.  It is known that for stationary ergodic processes, $\hat{HR}_{lz}(X)$ converges to the entropy rate $HR(X)$ with probability of $1$ as $n$ approaches infinity \cite{Kontoyiannis:1998}. It is important that in cases where the original data points are continuous (which is the case for financial markets) we need to discretize the data points for the purpose of the Lempel-Ziv complexity estimator. This procedure can be performed in many ways, the number of bins into which the data is assigned is a matter of convention and researchers choice, but it is advised that it should not be larger than square root of the sample size, and in fact should presumably be much smaller. In the case of financial markets we propose that the number of bins should be between 4 \cite{Fiedor:2013} and 8 \cite{Navet:2008}. It is important however that the states represent quartiles or other equal divisions, therefore each state is assigned the same number of data points. This design means that the model has no unnecessary parameters, which could affect the results and conclusions reached while using the data. This experimental setup also proved to be very efficient at revealing the randomness of the original data \cite{Steuer:2001}.

Based on this we can also define Lempel-Ziv complexity for multidimensional sequences. In fact the first attempt to use the Lempel-Ziv complexity for analysing spatio-temporal data has been presented over 25 years ago by Kaspar and Schuster \cite{Kaspar:1987}. A more natural approach extending the Lempel-Ziv complexity for vectorial data has been proposed in \cite{Zozor:2005}. This is done simply by extending the alphabet of the sequences. We consider $k$ sequences $X_{i} = x_{i,1}\ldots{}x_{i,n}$ for $i = 0,\ldots{}, k - 1$, where the letters are respectively in the alphabets $\mathcal{A}_{0},\ldots{},\mathcal{A}_{k-1}$ of sizes $\alpha_{0},\ldots{},\alpha_{k-1}$. Then we consider a sequence $Z = z_{1}\ldots{}z_{n}$ defined on the extended alphabet $\mathcal{B} = \mathcal{A}_{0} \times \; \ldots{} \times \mathcal{A}_{k-1}$ of size $\alpha_{0}\ldots{}\alpha_{k-1}$, the components of which are $k$-uplets $z_{j} = (x_{0,j},\ldots{},x_{k-1,j})$. $Z$ is then a sequence of $n$ $k$-uplets and not a sequence of $k \times{} n$ letters, therefore it does not result from a letter mixing approach. The approach defined by Lempel and Ziv holds for $k$-uplets, therefore all the work of Lempel and Ziv remains valid for vectorial sequences \cite{Lempel:1976}. Hence the joint Lempel-Ziv complexity of sequences $X_{0},\ldots{},X_{k-1}$ is defined as:
\begin{equation}
HR_{lz}(X_{0},\ldots{},X_{k-1}) = HR_{lz}(Z)
\end{equation}
Additionally if the alphabets are the same and are of the form $\mathcal{A} = \{0,\ldots{},\alpha{}-1\}$, we can also define sequence $Z = z_{1}\ldots{}z_{n}$ considering that each $z_j$ has the $x_{i,j}$ as $\alpha$-ary decomposition, that is $z_{j} = \sum_{i=0}^{k-1} x_{i,j}\alpha^{i}$. Defining joint Lempel-Ziv complexity of the $X_i$ as that of $Z$ is equivalent to the previous definition. Then Lempel-Ziv complexity of multidimensional sequences can then be viewed as a joint Lempel-Ziv complexity. 

Therefore, analogous with the Shannon information theory \cite{Shannon:1948}, mutual Lempel-Ziv complexity can be defined using the joint Lempel-Ziv complexity defined for two sequences $X$ and $Y$ as:
\begin{equation}
	\label{eq:MLZ_complexity}
	\hat{HR}_{lz}(X,Y) = \frac{n \log_2 n}{\sum_i \Lambda_i},
\end{equation}
where $i$ and $\lambda$ are defined over the joint sequence $Z$ defined above (as a union of $X$ and $Y$). Then mutual Lempel-Ziv complexity is defined as \cite{Blanc:2008}:
\begin{equation}
	\label{eq:MLZ_complexity2}
	MHR_{lz}(X,Y) = HR_{lz}(X) + HR_{lz}(Y) - HR_{lz}(X,Y).
\end{equation}
The mutual Lempel-Ziv complexity (MLZC) can be interpreted as a convergence measure between two sequences. Mutual Lempel-Ziv complexity can be negative transiently for finite $N$, but for $N \to \infty$ the asymptotic quantity $MHR_{lz}(X,Y)$ is always positive. In fact the MLZC converges asymptotically to a dynamic extension of the mutual information: the mutual information rate \cite{Ihara:1993,Pinsker:1964}.

We now know what mutual information rate is and how to estimate it asymptotically using Lempel-Ziv complexity for multidimensional data. But in order to create a topology of the dependence network we would prefer to have an Euclidean metric, which neither the mutual information nor mutual information rate are. Therefore we need to transform mutual information rate (equivalently mutual Lempel-Ziv complexity) into a measure which satisfies the axioms of an Euclidean metric. Here we will use the mutual information based metric proposed in \cite{Kraskov:2005}. Since mutual information and mutual information rate share most of their properties it's therefore possible to use this metric directly exchanging mutual information with mutual information rate.

Mutual information and mutual information rates are themselves similarity measures, although not well-defined ones, in the sense that small values imply large distances in a network. But it is useful to modify them such that the resulting quantity is a metric in the strict Euclidean sense. Indeed, the first such metric is well known \cite{Cover:1991}.
The quantity
\begin{equation}
d(X,Y)=H(X|Y)+H(Y|X)=H(X,Y)-I_{S}(X,Y)
\end{equation}
\begin{equation}
d(X,Y)=H(X)+H(Y)-2I_{S}(X,Y)
\end{equation}
satisfies the triangle inequality, is non-negative, symmetric and satisfies $d(X,X)=0$. This has been proved in \cite{Kraskov:2005}. But $d(X,Y)$ is not appropriate for all purposes. Since when constructing a network we may want to compare the proximity between two objects and two clusters of object (for example clustering sectors together), we would prefer the distance measure to be unbiased with regards to the size of the clusters. As argued forcefully in \cite{Li:2001} this is not true for $I_{S}(X,Y)$ or $d(X,Y)$. Mutual information depends on the size of the studied sequence.

Thus we form two different distances which measure relative distance, by being normalised through dividing by the total entropy. Then the quantity:
\begin{equation}
D(X,Y) = 1 - \frac{I_{S}(X,Y)}{H(X,Y)} = \frac{d(X,Y)}{H(X,Y)}
\end{equation}
is a metric, with $D(X,X)=0$ and $D(X,Y)\leq 1$ for all pairs $(X,Y)$.

Additionally the quantity:
\begin{eqnarray}
   D'(X,Y) & = & 1 - \frac{I_{S}(X,Y)}{\max\{H(X),H(Y)\}}    \nonumber \\
     & = & \frac{\max\{H(X|Y),H(Y|X)\}}{\max\{H(X),H(Y)\}}                \label{eq:dist2}
\end{eqnarray}
is also a metric, also with $D'(X,X)=0$ and $D'(X,Y)\leq 1$ for all pairs $(X,Y)$. It is sharper than $D$ in the sense that $D'(X,Y) \leq D(X,Y)$.

The practical advantage of $D'$ over $D$ has not been found \cite{Kraskov:2005} and therefore it's advisable to use $D$ due to its simpler nature. Since we are using mutual information rate we can use both $d$ and $D$ as the mutual information rate is defined per bit of information and therefore the drawbacks of $d$ defined over mutual information do not apply. The choice between $d$ and $D$ is thus less important with their being defined over mutual information rate, but we will use $D$ in this study for consistency.

Let us once again define $D$, this time in terms of mutual information rate:
\begin{equation}
 D(X,Y) = \frac{d(X,Y)}{HR(X,Y)},
\end{equation}
where:
\begin{equation}
d(X,Y)=HR(X,Y)-MIR(X,Y)
\end{equation}
\begin{equation}
d(X,Y)=HR(X)+HR(Y)-2MIR(X,Y)
\end{equation}

We now have a metric allowing us to quantify distance between nodes in hierarchical networks describing interdependencies on financial markets, therefore we can turn briefly to summarising the procedures used for creating such networks.

\section{Hierarchical Networks}

Having defined the distance measure we now briefly turn to the construction methods for two filtered graphs best-suited for financial market research, that is the minimal spanning tree and planar maximally filtered graph. These methods are well-known in literature, hence we will only briefly define them. The distance matrix $\mathcal{D}$ containing $D(X,Y)$ for all studied pairs is used to determine the minimal spanning tree and planar maximally filtered graph \cite{Papa:1982} connecting $n$ financial instruments in the studied set. On the basis of the distance matrix $D$ we create an ordered list $\mathcal{S}$, in which the distances are listed in decreasing order. Then, to create a minimal spanning tree, starting from the first element of the list the corresponding link is added to the network if and only if the resulting graph is still a forest or a tree \cite{Aste:2005}. Similarly a planar maximally filtered graph can be constructed in the same way by adding the corresponding link if and only if the resulting graph is still a planar graph (with genus equal $0$).

Such construction means that these methods filter significant information out of the characteristic vector describing the studied complex system, allowing the analyst to concentrate only on the most important information and dependencies within the system, facilitating the understanding of its behaviour. Similarity measure $\mathcal{D}$ is defined as a matrix of measure $D$ based on mutual information rate between all pairs of elements in the system, as opposed to the correlation coefficient matrix of the system used in other methodologies. An ordered list $\mathcal{S}$ is constructed by arranging them in descending order according to the value of the similarity $D(X,Y)$ between elements $X$ and $Y$. On this basis a number of filtering tools can be applies to create different network structures, here we first look at the method of constructing the minimal spanning tree (MST): using the ordered list $\mathcal{S}$ starting from the couple of elements with the largest similarity measure $D$ an edge is added to the graph between element $X$ and element $Y$ if and only if the graph obtained after such edge insertion is still a forest or a tree \cite{Aste:2005}. In fact when using this method the graph obtained after all appropriate links are added is reduced from a forest into a tree \cite{Aste:2005,Lillo:2010}.

Similarly to the above method for constructing the MST we can also construct graphs where the only topological constraint is the fixed genus $g = k$. Then the construction algorithm for these is also similar: again we follow the ordered list $\mathcal{S}$ starting from the couple of elements with the largest similarity, and we add an edge between that pair of elements on the list if and only if the resulting graph can still be embedded on a surface of genus $g \leq{} k$. Such graph embedded on a surface of genus $g = k$ is a simple, undirected, connected graph, which is less topologically restrictive than MST. In fact it's been proved that these graphs always contain the relevant MST and additionally they contain other relevant information associated with the structure of loops and cliques, making these graphs a natural extensions of the MST. A clique of $r$ elements ($r$-clique) is a complete subgraph that links all $r$ elements \cite{Aste:2005}.

When the genus is set to $g = 0$ then the resulting graph is planar \cite{Miller:1987}, that is it can is embedded on a sphere. Such graph is the natural extension of the MST, and is called the Planar Maximally Filtered Graph (PMFG). The fundamental difference between the PMFG and the MST is in the number of links: MST contains only $n - 1$ links while PMFG contains $3(n - 2)$ links. More generally the number of links in a graph with a genus $g = k$ is at most $3(n - 2 + 2k)$. The PMFG is the simplest graph extending the MST and the one providing only the most significant information in addition to the information already present in the MST. The topological constraints of the PMFG mean that each element of it has to participate in at least one clique of three elements. The PMFG is then a topological triangulation of the sphere \cite{Aste:2005}. Only cliques of three and four elements are present in the PMFG, as Kuratowski's theorem \cite{Miller:1987} does not allow cliques with a larger number of elements in a planar graph. Larger cliques can only be present in graphs with genus $g > 0$. The number of elements of the maximal allowed clique is growing with growing genus \cite{Ringel:1974}. 

The minimal spanning tree provides a very topologically restrictive arrangement of financial instruments, which selects the most relevant connections of each point of the set. Therefore the hierarchical organisation found this way is highly interesting from an economic point of view, providing the most important dependencies in the market, which has been shown in numerous studies. But when the structure of MST is too restrictive other structures proposed in literature besides minimal spanning tree, such as planar maximally filtered graph \cite{Lillo:2010} can be useful, and thus we will use both of those structures to illustrate the usage of mutual information rate as the similarity measure. It is worth noting that there are other structures when the mentioned two are not enough for a given application.

\section{Empirical Application}

To apply mutual information rate-based networks in practise we have taken log returns for 15 securities out of 20 which constitute the blue chip index of Warsaw's Stock Exchange (WIG20), that is all of the 20 which have been continuously traded during the studied period. The data has been downloaded from \url{http://bossa.pl/notowania/metastock/}, we have studied prices for years 2009-2012. The data is transformed in the standard way for analysing price movements, that is so that the data points are the log ratios between consecutive daily closing prices: $r_{t}=ln(p_{t}/p_{t-1})$ and those data points are, for the purpose of the Lempel-Ziv complexity estimator, discretized into 4 and 10 distinct states. The states represent equal parts, therefore each state is assigned the same number of data points. This design means that the model has no unnecessary parameters, which could affect the results and conclusions reached while using the data. This and similar experimental setups have been used in similar studies \cite{Navet:2008,Fiedor:2013} (Navet \& Chen divided data into 8 equal parts) and proved to be very efficient \cite{Steuer:2001,Navet:2008,Fiedor:2013}.

\begin{figure}[tbh]
\centering
\includegraphics[width=0.35\textwidth]{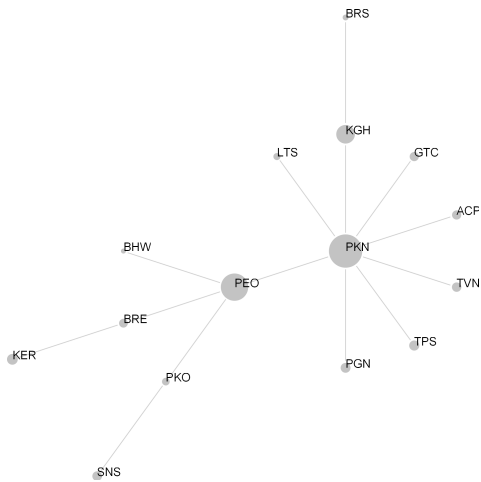}
\caption{MST based on correlation -- WIG20}
\label{fig:MST_corr}
\end{figure}

\begin{figure}[tbh]
\centering
\includegraphics[width=0.35\textwidth]{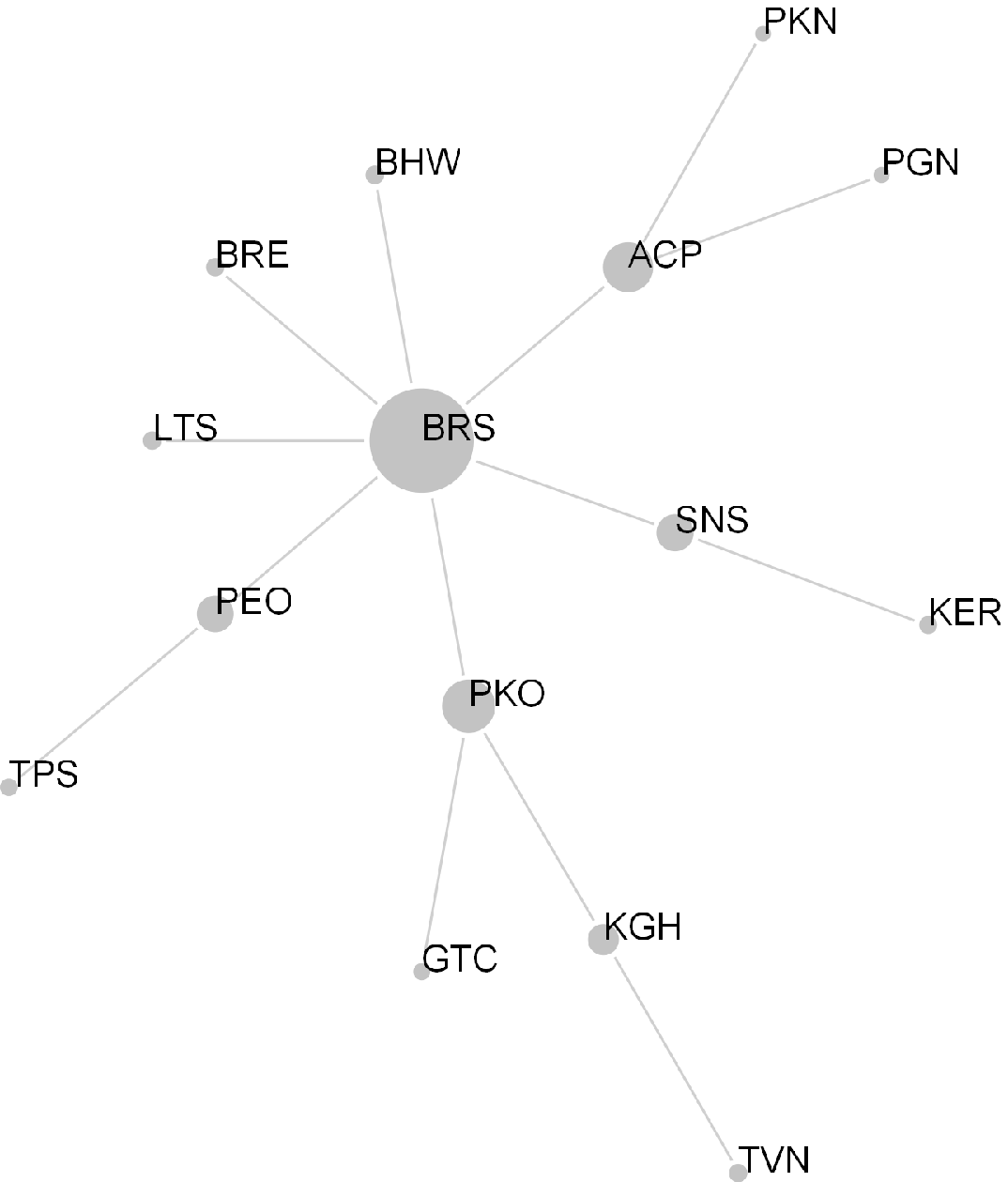}
\caption{MST based on MIR ($\alpha=4$) -- WIG20}
\label{fig:MST_bin_MI}
\end{figure}

\begin{figure}[tbh]
\centering
\includegraphics[width=0.3\textwidth]{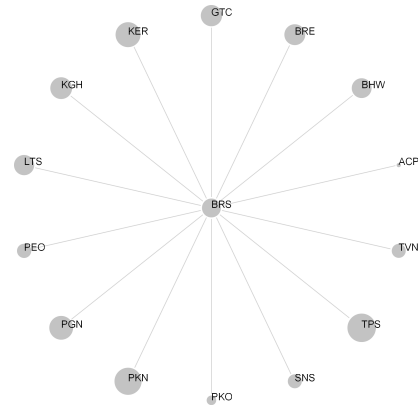}
\caption{MST based on MIR ($\alpha=10$) -- WIG20}
\label{fig:MTS_10_MI}
\end{figure}

We have therefore created three minimal spanning trees: for the original undiscretized log returns using correlation based distance (seen on Fig. \ref{fig:MST_corr}), and for discretized log returns with the alphabet of cardinality 4 and 10 using mutual information rate based distance (seen on Figs. \ref{fig:MST_bin_MI} \& \ref{fig:MTS_10_MI} respectively). The size of vertexes in all networks is dependent on Markov centrality.

\begin{figure}[tbh]
\centering
\includegraphics[width=0.35\textwidth]{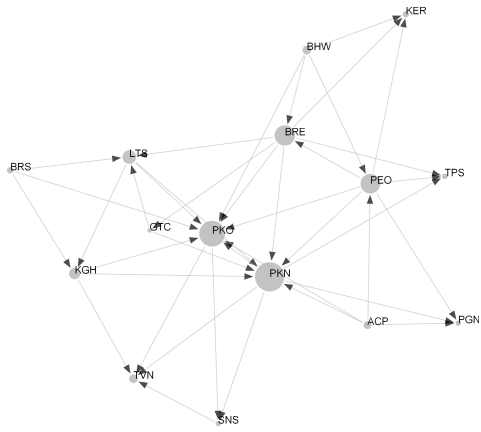}
\caption{PMFG based on correlation -- WIG20}
\label{fig:PMFG_corr}
\end{figure}

\begin{figure}[tbh]
\centering
\includegraphics[width=0.35\textwidth]{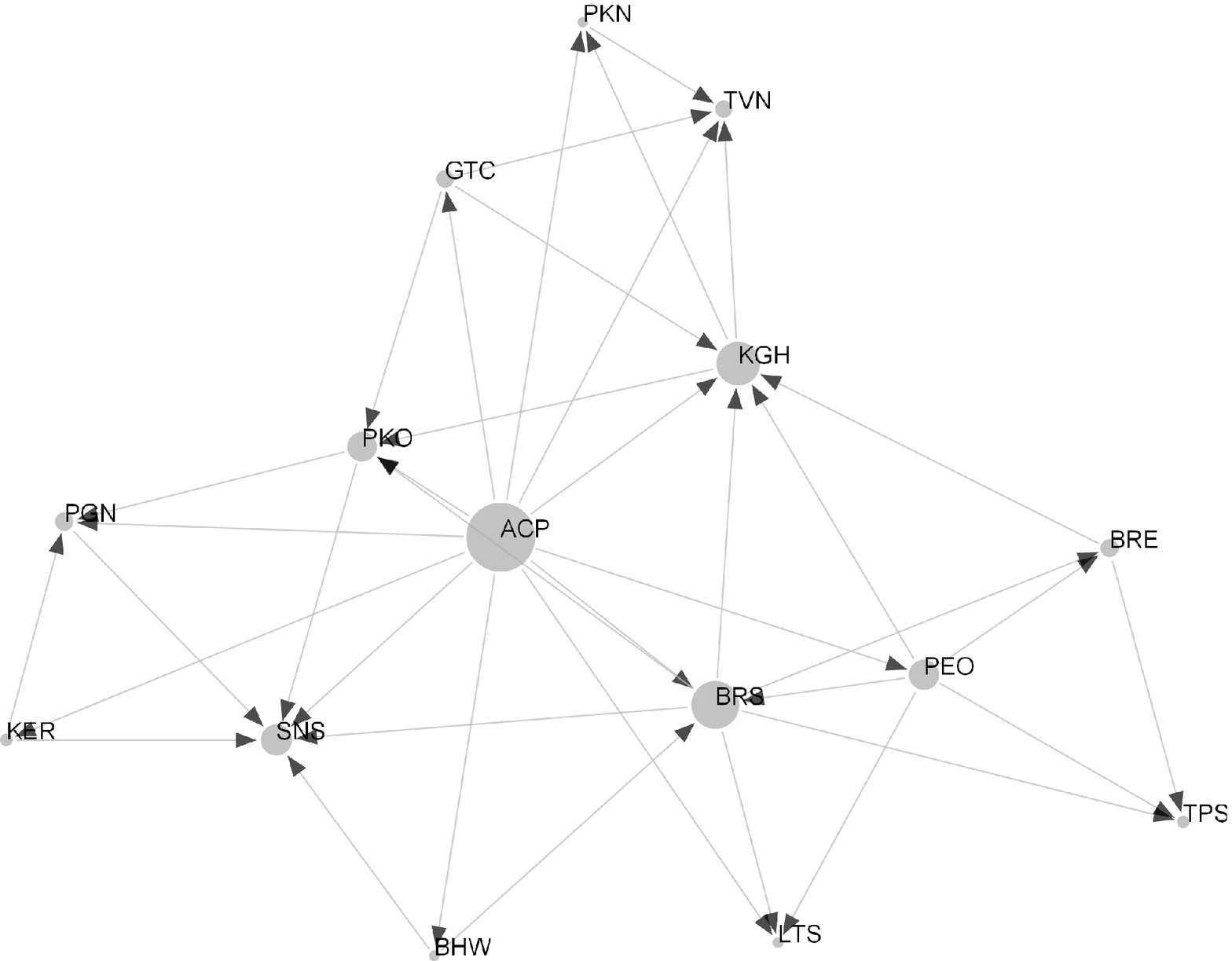}
\caption{PMFG based on MIR ($\alpha=4$) -- WIG20}
\label{fig:PMFG_bin_MI}
\end{figure}

\begin{figure}[tbh]
\centering
\includegraphics[width=0.35\textwidth]{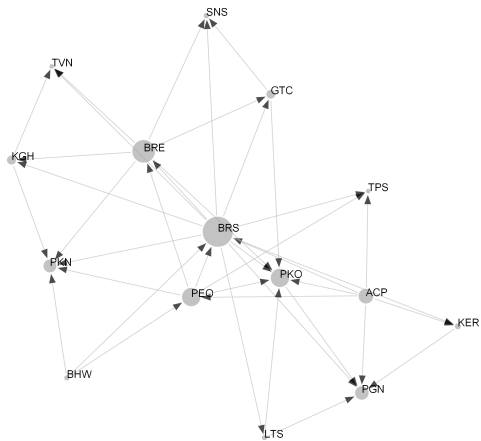}
\caption{PMFG based on MIR ($\alpha=10$) -- WIG20}
\label{fig:PMFG_10_MI}
\end{figure}

We have also created three planar maximally filtered graphs for the same experimental setup (seen on Figs. \ref{fig:PMFG_corr}, \ref{fig:PMFG_bin_MI} \& \ref{fig:PMFG_10_MI}). The choice of such small group of securities is based on the easily readable nature of smaller networks, therefore the changes between correlation and mutual information rate based networks are more intuitive graphically. Nonetheless we have also calculated Markov centrality measure for every node to quantitatively capture the differences between the networks as well.

\begin{figure}[tbh]
\centering
\includegraphics[width=0.35\textwidth]{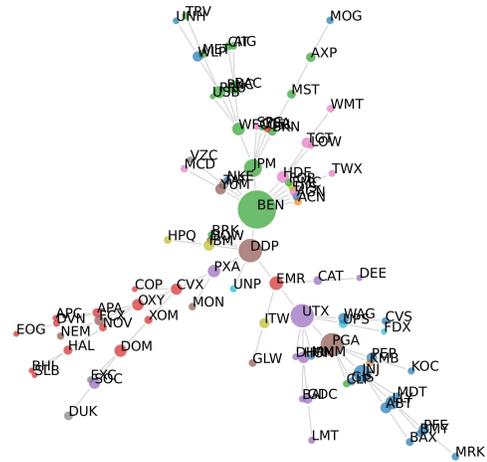}
\caption{MST based on correlation -- NYSE100 (Color online)}
\label{fig:MST_corr2}
\end{figure}

\begin{figure}[tbh]
\centering
\includegraphics[width=0.35\textwidth]{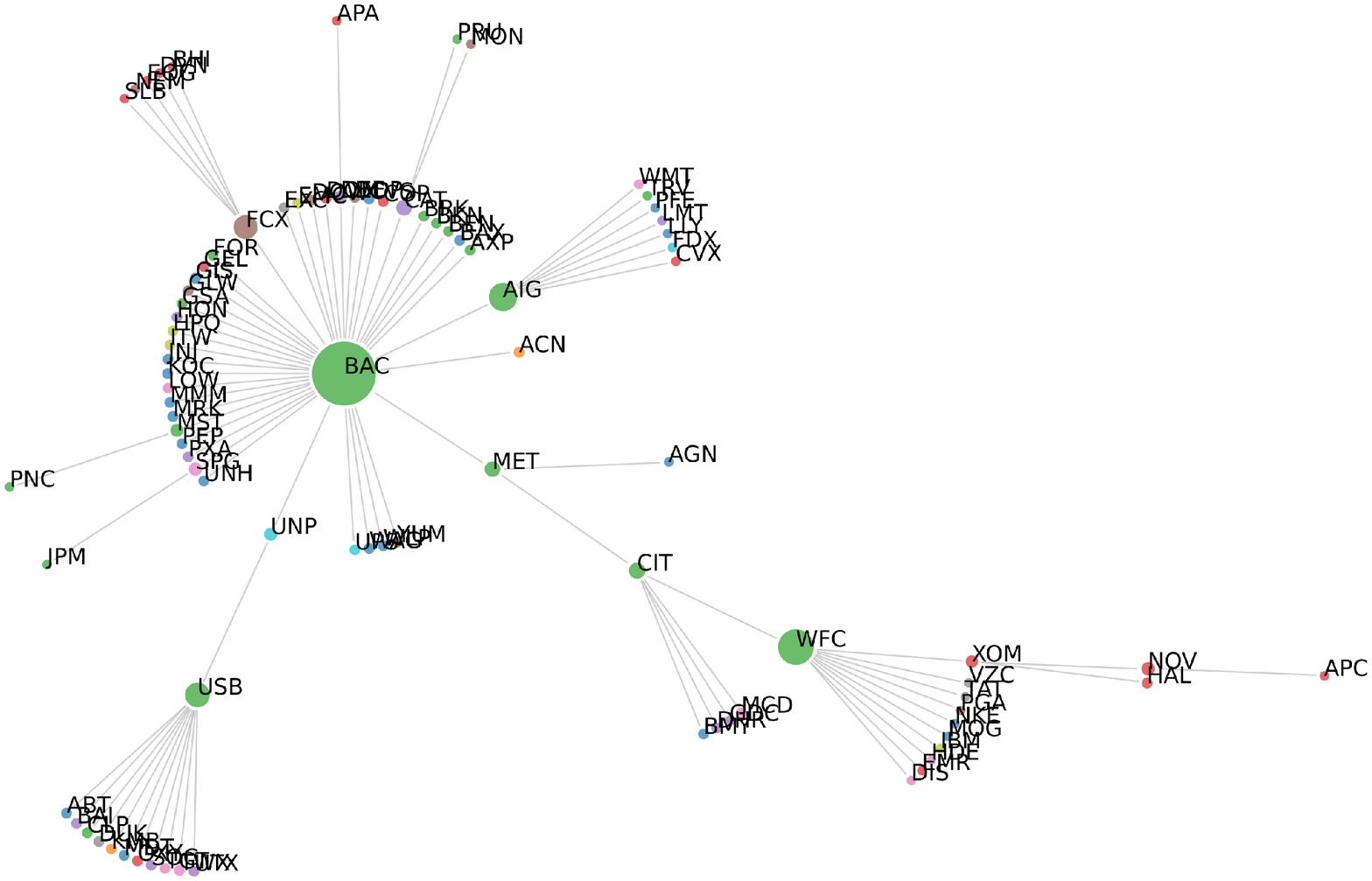}
\caption{MST based on MIR ($\alpha=4$) -- NYSE100 (Color online)}
\label{fig:MST_bin_MI2}
\end{figure}

\begin{figure}[tbh]
\centering
\includegraphics[width=0.3\textwidth]{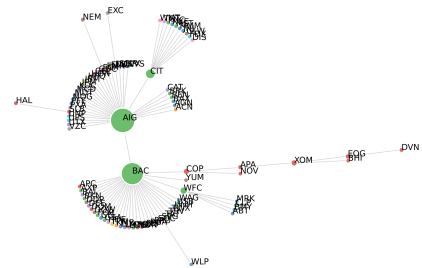}
\caption{MST based on MIR ($\alpha=10$) -- NYSE100 (Color online)}
\label{fig:MTS_10_MI2}
\end{figure}

Here we need to briefly define Markov centrality. One of the most important metrics in any network is centrality. Central nodes in a graph are often seen as the important agents, through which the interactions are conducted (be it social interactions, economic processes or biological interactions). Centrality is a good indicator of the relative popularity of individual nodes \cite{Faust:1994}. There are numerous ways to quantify centrality of a network \cite{Newman:2003}, their quality is usually comparable, hence we have chosen Markov centrality due to its general nature. Markov centrality interprets the network as a Markov process and can be intuitively understood as the amount of time a token performing a random walk spends on each node. This can be computed as the mean first-passage time in the Markov chain. For detailed description see \cite{Jeong:2001,White:2003,Brandes:2005}.

Markov centralities for nodes in the studied minimal spanning trees are presented in Table \ref{tab:mst_mc} and for the studied planar maximally filtered graphs in Table \ref{tab:pmfg_mc}.

\begin{table}[htbp]
%\centering
\caption{Markov centrality for MST}
\begin{ruledtabular}
\begin{tabular}{lrrr}
%\addlinespace
%\hline
Vertex & MSTcorr & MSTmir4 & MSTmir10 \\ \hline
ACP & 0.050 & 0.188 & 0.065 \\
BHW & 0.025 & 0.048 & 0.067 \\
BRE & 0.047 & 0.048 & 0.067 \\
BRS & 0.030 & 0.237 & 0.067 \\
GTC & 0.049 & 0.049 & 0.067 \\
KER & 0.059 & 0.025 & 0.067 \\
KGH & 0.105 & 0.047 & 0.067 \\
LTS & 0.037 & 0.048 & 0.067 \\
PEO & 0.155 & 0.047 & 0.066 \\
PGN & 0.053 & 0.024 & 0.067 \\
PKN & 0.190 & 0.024 & 0.068 \\
PKO & 0.041 & 0.047 & 0.066 \\
SNS & 0.052 & 0.096 & 0.066 \\
TPS & 0.055 & 0.049 & 0.067 \\
TVN & 0.051 & 0.024 & 0.066 \\
\end{tabular}
\end{ruledtabular}
\label{tab:mst_mc}
\end{table}

\begin{table}[htbp]
%\centering
\caption{Markov centrality for PMFG}
\begin{ruledtabular}
\begin{tabular}{lrrr}
%\addlinespace
%\hline
Vertex & PMFGcorr & PMFGmir4 & PMFGmir10 \\ \hline
ACP & 0.052 & 0.165 & 0.103 \\
BHW & 0.055 & 0.052 & 0.039 \\
BRE & 0.099 & 0.038 & 0.117 \\
BRS & 0.046 & 0.164 & 0.171 \\
GTC & 0.038 & 0.052 & 0.052 \\
KER & 0.045 & 0.040 & 0.039 \\
KGH & 0.064 & 0.051 & 0.052 \\
LTS & 0.073 & 0.039 & 0.039 \\
PEO & 0.096 & 0.064 & 0.077 \\
PGN & 0.042 & 0.051 & 0.052 \\
PKN & 0.132 & 0.052 & 0.039 \\
PKO & 0.119 & 0.064 & 0.064 \\
SNS & 0.042 & 0.077 & 0.039 \\
TPS & 0.043 & 0.040 & 0.052 \\
TVN & 0.054 & 0.051 & 0.065 \\
\end{tabular}
\end{ruledtabular}
\label{tab:pmfg_mc}
\end{table}

\begin{figure}[tbh]
\centering
\includegraphics[width=0.35\textwidth]{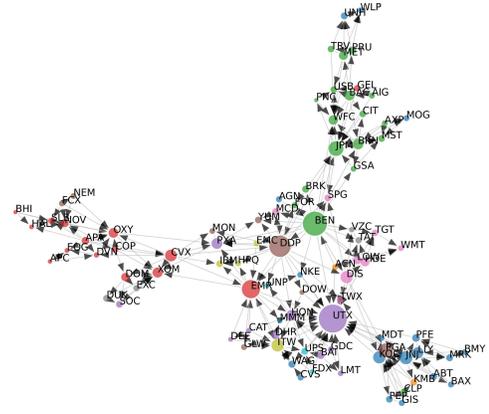}
\caption{PMFG based on correlation -- NYSE100 (Color online)}
\label{fig:PMFG_corr2}
\end{figure}

Correlations between MST based on correlation and MST based on mutual information rate discretized into 4 and 10 states are equal to $-0.19$ and $0.22$ respectively. Equivalent correlations for planar maximally filtered graphs are equal to $-0.14$ and $0.36$ respectively. This and the resulting graphical representations show that the networks are not containing the same information. Thus using mutual information rate instead of correlation alters the analysis. Since mutual information and mutual information rate measure both the linear and non-linear dependencies we believe that the analysis using these measures is interesting and further studies should be performed to see whether it is useful in analysing other markets. Nonetheless the discretisation step appears to be important and the choice of the number of bins appears relevant. There is no guideline to this, we only advise the upper limit to be well below the square root of the sample size, otherwise the choice is dependent on what the research wants to study (more or less granular price changes). It is worth noting that we believe in our example alphabet cardinality of 4 performs better. Another issue with using Lempel-Ziv complexity is that it's only asymptotically equal to mutual information rate, therefore we don't recommend using this method for small samples, that is ones with cardinality of under 500 data points (which amount to about 2 years for daily financial data).

\begin{figure}[tbh]
\centering
\includegraphics[width=0.35\textwidth]{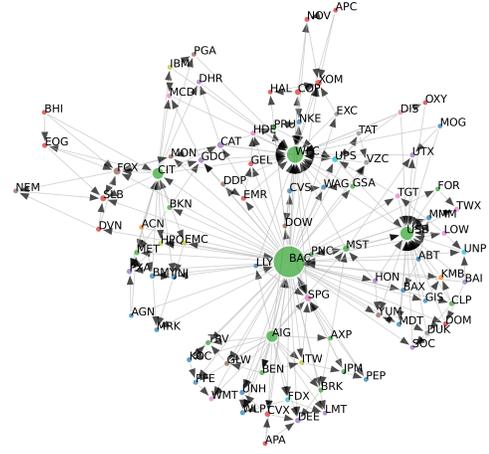}
\caption{PMFG based on MIR ($\alpha=4$) -- NYSE100 (Color online)}
\label{fig:PMFG_bin_MI2}
\end{figure}

\begin{figure}[tbh]
\centering
\includegraphics[width=0.3\textwidth]{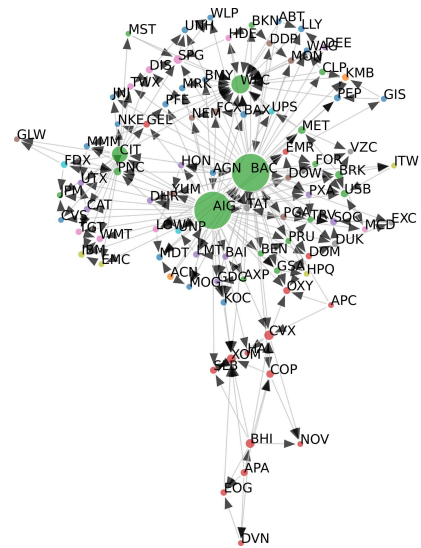}
\caption{PMFG based on MIR ($\alpha=10$) -- NYSE100 (Color online)}
\label{fig:PMFG_10_MI2}
\end{figure}

Recognising that a sample of 15 stocks in a period of three years is somewhat limiting to the analysis despite the illustratory power we have also calculated the same for 91 stocks belonging to NYSE100 index which were traded continuously between 11th of November 2003 and 7th of November 2013. Correlations between MST based on correlation and MST based on mutual information rate discretized into 4 and 10 states are equal to $0.05$ and $0.01$ respectively. Equivalent correlations for planar maximally filtered graphs are equal to $0.03$ and $0$ respectively. The resultings spanning tree for the original undiscretized log returns using correlation based distance can be seen on Fig. \ref{fig:MST_corr2}, and for discretized log returns with the alphabet of cardinality 4 and 10 using mutual information rate based distance can be seen on Figs. \ref{fig:MST_bin_MI2} \& \ref{fig:MTS_10_MI2} respectively. We have also created three planar maximally filtered graphs for the same experimental setup (seen on Figs. \ref{fig:PMFG_corr2}, \ref{fig:PMFG_bin_MI2} \& \ref{fig:PMFG_10_MI2}).

\section{Conclusions}

In this paper we have presented a methodology for creating hierarchical networks studying financial markets using dynamical extension of mutual information called mutual information rate, which we estimate using multidimensional Lempel-Ziv complexity. We have applied this methodology to Warsaw and New York stock exchanges (WIG20 and NYSE100). The resulting minimal spanning tress and planar maximally filtered graphs are significantly different from those obtained using Pearson's correlation as similarity measure, therefore we conclude that the non-linear dependencies not captured by Pearson's correlation coefficient and captured by mutual information rate are indeed relevant to the hierarchical structure of the financial markets. The proposed methodology is sensitive to the choice of number of bins into which the log returns are discretized, and requires large sample sizes. Further research should look into the differences between networks based on mutual information and mutual information rate, and also into other estimators of these measures tailer for hierarchical clustering, as well as the determination of the best way to discretize log returns for the purpose of hierarchical clustering using information theoretic approach. Further studies based on other stock markets, market indexes and currency exchange markets should also be performed to analyse the usefulness of this approach.

\bibliography{prace}
\end{document}